\documentclass[showpacs,preprintnumbers,amsmath,amssymb]{revtex4}
\usepackage{graphicx}
\usepackage{epsfig}
\usepackage{bbm}
\begin{document}

\title{Efimov spectrum in bosonic systems with increasing number of
   particles }

\author{A. Kievsky}
\affiliation{Istituto Nazionale di Fisica Nucleare,
              Largo Pontecorvo 3, 56127 Pisa, Italy}
\author{M. Gattobigio}
\affiliation{Universit\'e de Nice-Sophia Antipolis, Institut Non-Lin\'eaire de
Nice,  CNRS, 1361 route des Lucioles, 06560 Valbonne, France}
\author{N.K. Timofeyuk }
\affiliation{Department of Physics, University of Surrey, Guildford, Surrey GU2 7XH, United Kingdom
}

\begin{abstract}

It is well-known that three-boson systems show the Efimov effect
when the two-body
scattering length $a$ is large with respect to the range of the two-body interaction.
This effect is a manifestation of a discrete scaling invariance (DSI).
In this work we study DSI in the $N$-body system by analysing 
the spectrum of $N$ identical bosons obtained with a pairwise gaussian interaction 
close to  the unitary limit.  We consider different universal ratios such as
$E_N^0/E_3^0$ and $E_N^1/E_N^0$, with $E_N^i$ being the energy of the ground 
($i=0$) and first-excited ($i=1$) state of the system, for  $N\le16$. 
We discuss the extension of the Efimov radial law,
derived by Efimov for $N=3$, to general $N$.
\end{abstract}

\maketitle

\section{Introduction}
\label{intro}
In the unitary limit, when the two-body scattering length $a$ of two identical bosons
goes to $\pm\infty$, the three-boson spectrum has an infinite set of bound states,
$E_3^n$, approaching zero in a geometrical progression. This is called Efimov effect 
and is related to a discrete scaling invariance (DSI) in the system of three identical 
bosons with total angular momentum $L=0$. Explicitly,
the ratio $E_3^{n+1}/E_3^n$ is equal to ${\rm e}^{-2\pi/s_0}\approx 1/515.03$ with
$s_0$ an universal number. As the
absolute value of $a$ takes finite values, the three-body highest bound states
disappear either into the atom-dimer continuum ($a>0$) or in the
three-atom continuum ($a<0$). In recent years, the spectrum of the
three-boson system has been extensively studied in the
$(1/a,\kappa)$ plane, where $\kappa^2=mE/\hbar^2$ \cite{report}.
When one boson is added to the three-boson system, the resulting four-body system at the
unitary limit has two bound states, one deep ($E_4^0$) and one
shallow ($E_4^1$) with the ratios $E_4^0/E_3^0\approx 4.6$
and $E_4^1/E_3^0\approx 1.001$, having a  universal character \cite{deltuva1}.
This particular form of the spectrum has been recently studied up to six
bosons \cite{gatto2012}.

In the present work we investigate the spectrum of $N\le16$ bosons.
We compute different universal ratios,
$E_N^0/E_3^0$ and $E_N^1/E_N^0$, close to the unitary limit to
study the consequence of the three-boson DSI in the $N$-body system.
In particular we extract some relations for $E_N^0\rightarrow 0$.

\section{$N$-boson energy spectrum}
\label{sec:1}

Following Ref.~\cite{gatto2012}, we
describe the $N$-boson system using a two-body gaussian (TBG) potential
\begin{equation}
 V(r)=V_0 {\rm e}^{-r^2/r_0^2} \,\, ,
\end{equation}
and we solve the Schr\"odinger equation with mass parameter $\hbar^2/m=43.281307$
$(a_0)^2$K. With $r_0=10\, a_0$ and $V_0=-1.2343566\,$K the model reproduces
the binding energy and the scattering length of two helium atoms described
by a widely used He-He interaction, the LM2M2
potential~\cite{lm2m2}. Varying the strength $V_0$ of the potential the
$(a^{-1},\kappa)$ plane can be explored.

\begin{center}
  \begin{figure}[h]
    \includegraphics[width=10cm,height=8cm]{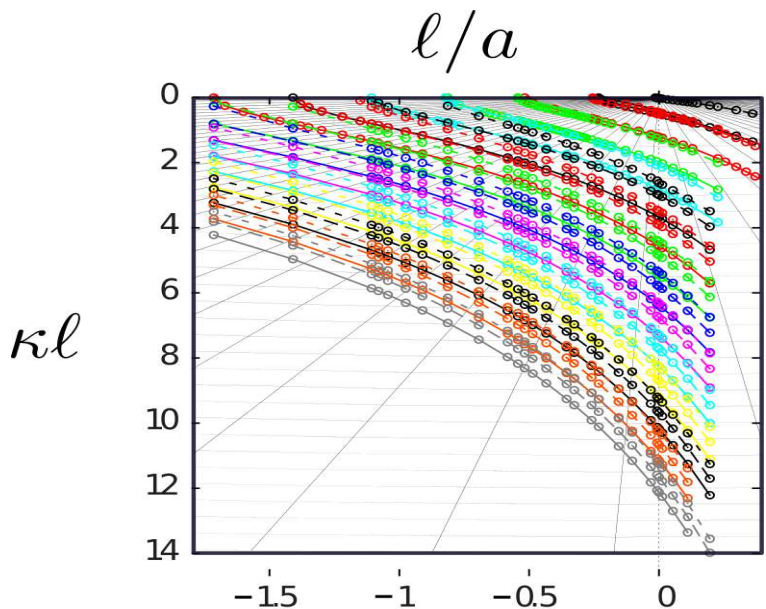}
    \caption{(Color on line) The $N=3-16$ ground state (solid lines) and
 excited state (dashed line) energy wave numbers (in units of the van der Waals
length $\ell$) as a function of
$\ell/a$. For $N\ge7$ the second excited state is also shown (dotted line).
The states organized progressively starting with the $N=3$ (black line on the
top) up to the $N=16$ (grey line on the bottom). For the sake of
illustration, lines at constant values of $\xi$ and circles at constant values
of $H$ are displayed.}
  \label{fig:polar}
  \end{figure}
\end{center}

\begin{center}
  \begin{figure}[h]
    \includegraphics[width=10cm,height=8cm]{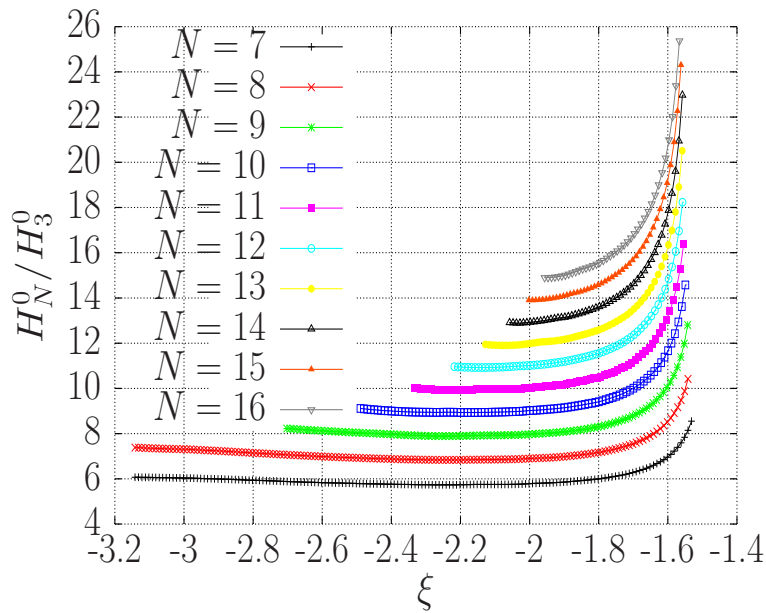}
    \caption{Ratios of the $N$-boson ground state polar coordinate $H_N^0$
 with respect  to the three-boson coordinate $H_3^0$ as a function of the
 angle $\xi$}
  \label{fig:ratios0}
  \end{figure}
\end{center}

To solve the Schr\"odinger equation for $N$ bosons we use the Hyperspherical
Harmonic (HH) method in the version proposed in Ref.\cite{natasha}. This method
reproduces the values given in Ref.~\cite{gatto2010} up to $N=6$ and 
it gives improved convergence with increasing $N$ as discussed in \cite{Tim12}. Here we
extend the HH calculations up to $N=16$. A first presentation of our results
is given in Fig.~\ref{fig:polar}.
We show the $N$-boson spectrum in the form of the energy wave number
$\kappa_N={\rm sign}(E_N)[|E_N|/(\hbar^2/m)]^{1/2}]$ given in terms of the inverse
value of the two-body scattering length $a$, both expressed in units of the van 
der Waals length $\ell=10.2\,$a.u., calculated for the LM2M2 potential~\cite{report}.
The ground state wave numbers are given by solid lines
starting from $N=3$ (black solid line) up to $N=16$ (gray solid
line). The excited state wave numbers are given by the same colored dashed lines
and in most cases this state is bound and close to the $N-1$ ground state.
Starting at $N=7$ the second excited state is shown by the colored dotted lines.
In the explored region, this state is bound for large positive values of
$a$ for  $N\ge 10$. As $a$ becomes negative this state becomes
progressively unbound and at values of $\xi\approx -95^\circ$ (see below)
it disappears for all the studied $N$. 
Defining the polar coordinates
\begin{eqnarray}
  H^2_N &=& 1/a^2 + E_N/(\hbar^2/m) \\
  \xi &=& \arctan (E_N/(\hbar^2/ma^2)) \,,
  \label{eq:polar_N}
\end{eqnarray}
which in the $N=3$ case reduce to the polar coordinates introduced by Efimov, 
we plot in Fig. 1 the lines at constant values of $\xi$
and circles at constant values of $H$  for eye guidance. The
DSI manifests itself via constant values between energy ratios at fixed values
of the angle $\xi$. In order to study DSI in the $N$-boson spectrum,
 we show the ratios between the ground state
polar coordinate $H_N^0$ and the three-body ground state coordinate
$H_3^0$ in Fig.~\ref{fig:ratios0}. A clear trend of constant behavior can be observed
as $a$ becomes negative
($\xi<-\pi/2$).

The radial law, derived by V. Efimov for  three-boson systems~\cite{efimov1}, 
gives the energy spectrum as a function of $a$. From our results
(see also Ref.~\cite{gatto2013}) we propose
the extension of the radial law to
general values of $N$, which in the limit of the $N$-boson ground
state energies,  is given by the equation
\begin{equation}
 E^0_N+\frac{\hbar^2}{ma^2}=\exp{[\Delta(\xi)/s_0]}\frac{\hbar^2(\kappa^0_N)^2}{m}\,.
\label{eq:uniN}
\end{equation}
Here $\kappa^0_N$ is the wave number corresponding to the energy
at the resonant limit and $\Delta(\xi)$ an universal function.
The parametrization of the universal function $\Delta(\xi)$ has been determined using effective field
theory  and it can be found in Ref.~\cite{report}. Fixing
the angle $\xi$, the values of $a$ in the above equation are
different for each $N$. Accordingly experimental studies can be
done in systems in which the scattering length
can be modified as in trapped atoms using Fesbach resonances 
(see for example Ref.~\cite{ferlaino}).

\begin{center}
  \begin{figure}[h]
    \includegraphics[width=10cm,height=8cm]{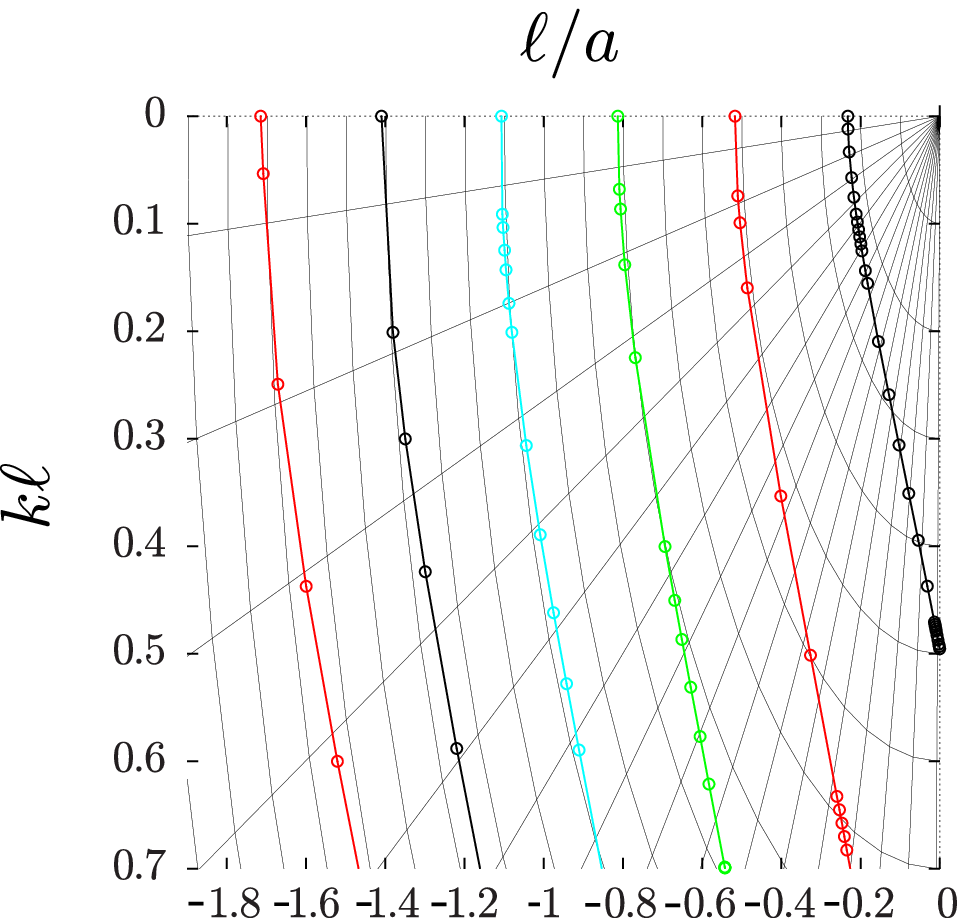}
    \caption{The energy wave number, in units of $\ell$, close
to threshold for $N=8$ (red line) up to $N=3$ (black line)}
  \label{fig:thresholds}
  \end{figure}
\end{center}

A second remark deduced from our results regards
the behavior of the energy curves  for $N=3-8$ approaching
the $\xi=-\pi$ axis as shown in Fig.~\ref{fig:thresholds}.
The quantity $H^0_N=\kappa_N\cos\xi$ is almost constant as the clusters
reach the $N$-body continuum and, as a consequence, the energy wave number follows
a circle line. Moreover, on the axis, the difference
$|\ell/a_N^-| - |\ell/a_{N-1}^-|$
is almost constant in all cases (a similar observation has been made in
Ref.~\cite{stecher}, see also Ref.~\cite{fedorov}). 
Applying Eq.(\ref{eq:uniN}) to $\xi=-\pi$
we obtain
\begin{equation}
 \kappa^0_Na_N^-=\exp{[-\Delta(-\pi)/s_0]}\approx-1.56(5) \,\, ,
\label{eq:unia}
\end{equation}
where we have used the prediction of Ref.~\cite{report}. The above
equation is a generalization of the $N=3$ case for a general $N$. The observed constant
difference between the $a_N^-$ together with Eq.(\ref{eq:unia}) can be
used to propose the following universal linear behavior
\begin{equation}
 \frac{\kappa^0_N}{\kappa^0_3}= 1+\gamma(N-3) \,\, ,
\label{eq:unik}
\end{equation}
with $\gamma=(\kappa^0_4/\kappa^0_3)-1$ an universal constant. The ratio
$\kappa^0_4/\kappa^0_3$ can be extracted, for example, from Ref.~\cite{deltuva1}
resulting in $\kappa^0_N/\kappa^0_3= 1+1.147(N-3)$. The square of this
relation gives a quadratic dependence of the energy with $N$ at the unitary
limit which has been already observed in Ref.~\cite{nicholson} though with
different constants.

\vspace{-3cm}
\section{Conclusions}
\label{sec:2}

In the present work, we have shown the spectrum of a system of $N$ 
identical bosons obtained with a fixed-range gaussian
potential. Varying its strength, we have explored 
an extended range of values of the two-body scattering length including
the unitary limit, $a\rightarrow\infty$. In particular, we have explored
the $\xi=-\pi$ axis where the $N$-body clusters disappear into the
$N$-body continuum and we have computed various ratios at fixed values of the
angle $\xi$ in order to study DSI. The main results of this analysis
is the extension of the Efimov radial law to general $N$, given in
Eq.(\ref{eq:uniN}), and the linear relation with $N$ between
$\kappa^0_N$ and the three-body parameter $\kappa^0_3$. Further studies
of this subject are in progress.

 \begin{acknowledgements}
 One of the authors, N.K.T., ackhowledges UK STFC Grant No. ST/J000051/1
 \end{acknowledgements}


\begin{thebibliography}{3}
\bibitem{report}
Braaten, E. and Hammer, H.: Universality in few-body systems with large
  scattering length.
\newblock Physics Reports \textbf{428}, 259 (2006)
\bibitem{deltuva1}
Deltuva, A., Lazauskas, R., and Platter, L.: Universality in {Four-Body}
  Scattering.
\newblock {Few-Body} Syst. \textbf{51}, 235 (2011)
\bibitem{gatto2012}
Gattobigio, M., Kievsky, A., and Viviani, M.: Energy spectra of small bosonic
  clusters having a large two-body scattering length.
  \newblock Phys. Rev. A \textbf{86}, 042513 (2012)
\bibitem{lm2m2}
Aziz, R.A. and Slaman, M.J.: An examination of ab initio results for the helium
  potential energy curve.
\newblock J. Chem. Phys. \textbf{94}, 8047 (1991)
\bibitem{natasha}
Timofeyuk, N. K.: Improved procedure to construct a hyperspherical basis for the N-body problem: Application to bosonic systems, 
  \newblock Phys. Rev. C \textbf{78},  054314  (2008)
\bibitem{gatto2010}
Gattobigio, M., Kievsky, A., and Viviani, M.: Spectra of helium clusters with
  up to six atoms using soft-core potentials.
\newblock Phys. Rev. A \textbf{84}, 052503 (2011)
\bibitem{Tim12}
Timofeyuk, N. K.: Convergence of the hyperspherical-harmonics expansion with increasing number of particles for bosonic systems,
  \newblock Phys. Rev. A \textbf{86},  032507 (2012)
\bibitem{efimov1}
Efimov, V.: Energy levels arising from resonant two-body forces in a three-body
  system.
\newblock Phys. Lett. B \textbf{33}, 563 (1970)
\bibitem{gatto2013}
Gattobigio, M. and Kievsky, A.: Universality and scaling in the $N$-body
sector of Efimov physics
\newblock submitted to publication, [arXiv:cond-mat/1309.1927]
\bibitem{ferlaino}
Ferlaino, F. and Grimm, R.: Forty years of Efimov physics: How a bizarre
  prediction turned into a hot topic.
\newblock Physics \textbf{3}, 9 (2010)
\bibitem{stecher}
von Stecher, J.: Weakly bound cluster states of Efimov character.
\newblock J. Phys. B: At. Mol. Opt. Phys. \textbf{43}, 101002 (2010)
\bibitem{fedorov}
Th\o gersen M., Fedorov D.V., and Jensen A.S.:
N-body Efimov states of trapped bosons
\newblock EPL, \textbf{83} (2008) 30012
\bibitem{nicholson}
Nicholson, Amy N.: $N$-body Efimov states from two-particle noise.
\newblock Phys. Rev. Lett. \textbf{109}, 073003 (2012)
\end{thebibliography}
\end{document}